\documentclass[%
 reprint,
 amsmath,amssymb,
 aps,
]{revtex4-1}

\usepackage{graphicx}
\usepackage{dcolumn}
\usepackage{bm}

\begin{document}

\preprint{APS/123-QED}


\title{Necessary and sufficient conditions for local creation of quantum discord\\}

\author{Yu Guo}
\altaffiliation{guoyu3@yahoo.com.cn}
\affiliation{Department of Mathematics, Shanxi Datong University, Datong 037009, China}
\affiliation{Institute of Optoelectroinc Engineering, Taiyuan University of Technology, Taiyuan 030024, China}

\author{Jinchuan Hou}
\altaffiliation{jinchuanhou@yahoo.com.cn}
\affiliation{{Department of Mathematics, Taiyuan University of Technology, Taiyuan 030024, China}}

\date{ \today}

\begin{abstract}

We show that a local channel cannot create quantum discord (QD)
for zero QD states of size $d\geq3$ if and only if either it is a
completely decohering channel or it is a nontrivial isotropic channel.
For the qubit case this propertiy is additionally characteristic to
the completely decohering channel or the commutativity-preserving unital channel.
In particular, the exact forms of the completely decohering channel and the commutativity-preserving unital qubit channel are proposed.
Consequently, our results confirm and improve the conjecture proposed by X.~Hu \emph{et al}. for the case of $d\geq3$ and improve the result proposed by
A.~Streltsov \emph{et al}. for the qubit case.
Furthermore, it is shown that a local channel nullifies QD in any state
if and only if it is a completely decohering channel.
Based on our results, some protocols of quantum
information processing issues associated with QD,
especially for the qubit case,
would be experimentally accessible.

\if
In this paper, we confirm the conjecture proposed in [X.~Hu \emph{et al}., Phys. Rev. A \textbf{85}, 032102(2012)]:
A local channel cannot create quantum discord (QD)
for zero QD states if and only if it is either a
completely decohering channel or an isotropic channel
for any qudit systems with $d\geq3$. In addition, the
explicit form of a local channel that cannot create QD
for zero QD states is presented for the qubit system.
This improves the result proposed by
A. Streltsov \emph{et al}. \if , Phys. Rev. Lett. \textbf{107}, 190502(2011)].\fi
Moreover, we show that i) a qudit ($d\geq3$)
local channel preserves zero QD states in both
directions (namely, the output state under the
channel has zero QD if and only if the input
state has zero QD) if and only if it is a nontrivial
isotropic channel, ii) the qubit case is far different
from the higher dimensional case: there exist non
isotropic local channels that preserve
zero QD states in both directions for the qubit system,
and iii) a local channel nullifies QD in any state
if and only if it is a completely decohering channel.
Several methods of testing whether
a given local channel can create QD are investigated.
In particular, the explicit form of completely decohering
channel is investigated.
Based on our results, some protocols of quantum
information processing issues associated with QD,
especially for the qubit case,
would be experimentally accessible.
\fi
\begin{description}
\item[PACS numbers]03.65.Ud, 03.65.Db, 03.65.Yz.
\end{description}
\end{abstract}

\pacs{Valid PACS appear here}
\maketitle

\section{Introduction}

The characterization of quantum correlated composite quantum systems
is an important topic in quantum information theory
\cite{Nielsen,Horodecki1,Guhne,Ollivier,Henderson,Luo1}. Different
approaches for characterizing the quantum correlations were studied in
the past two decades
\cite{Horodecki1,Guhne,Ollivier,Henderson,Luo1,Modi,Lewenstein,Chen,
Rudolph,Guo1,Horodecki3,Stormer,Xizhengjun,Wangzizhu,Augusiak,
Wushengjun,Gessner,Siewert,Hou2,Hou3,Luoshun,Hou}.
Recently, much interest has been devoted to the study of quantum
correlations that may arise without entanglement, such as quantum
discord (QD) \cite{Ollivier}, measurement-induced nonlocality
(MIN)\cite{Luo1} and quantum deficit \cite{Oppenheim}, etc. These
quantum correlations can still be a resource for a number of quantum
information applications \cite{Luo1,Modi,Dakic,Qasimi,Roa}. So,
this makes it important to understand the dynamics of these quantum
correlations under local noises (or operations) better.

In the qubit case, Streltsov \emph{et al}. showed in Ref.
\cite{Streltsov} that qubit channel that preserves
commutativity is either unital, i.e., mapping maximal mixed state to
maximal mixed state, or a completely decohering channel, i.e.,
nullifying QD in any state. In Ref.~\cite{Hu}, for $m\otimes3$ system, it
is showed that a channel $\Lambda$ acting on the second subsystem
cannot create QD for zero
QD states if and only if $\Lambda$ is either a completely decohering
channel or an isotropic channel. And it is  conjectured
that this result is also valid for any $m\otimes n$ system with
$n\geq3$. In Refs.~\cite{Hu,Yusixia}, the authors proved that, for any
$m\otimes n$ system with $mn<+\infty$, a channel $\Lambda$
(acting on the second subsystem)
transforms the zero QD states to zero QD states if and only if
$\Lambda$ preserves commutativity.
The goal of this paper is to propose an explicit
form of commutativity-preserving channel, from which we will i)
give a positive answer to
the above conjecture raised in Ref.~\cite{Hu}, ii) propose an exact
form of commutativity-preserving unital channel for the qubit system, and iii)
present an explicit form of
`completely decohering channel' for any system.

Beside fundamental interest, our results may result in useful applications.
The concrete form of commutativity-preserving unital qubit channel
may lead to a number of experimental tasks based on QD
since the qubit state is a direct resource in
various quantum information processing.
From our results, we know exactly whether a local
channel can create or nullify quantum correlation measured by QD.
We will show that we have more choices of local
channels for qubit case in the issue of preserving zero QD in the state:
there exist non-isotropic channels which also cannot
create QD for qubit case while only the isotropic channels
cannot create QD for higher dimensional case.
In addition, for both qubit and the higher dimension
cases, we provide several equivalent methods of determining whether
a given local channel can create QD.

The paper is organized as follows. In Sec.~II we
review the definitions of quantum channel
and QD, and fix some terminology. In Sec.~III we discuss
the local channels that cannot create QD for zero QD states,
and then in Sec.~IV, we deal
with the the local channels that nullifies QD in any states.
A summarization and some related
questions are listed at the end.

\section{Definitions and terminologies}

\if For clarity, we review the definitions of channel and QD,
respectively.\fi
Let $H$ be a complex Hilbert
space describing a quantum
system, $\dim
H=n<+\infty$. Let $\mathcal{B}(H)$ be the space of all
linear operators on $H$, and $\mathcal{S}(H)$ the set consisting
of all quantum states acting on $H$. Recall that,
a quantum channel (or channel, briefly) is described
by a trace-preserving completely
positive linear map $\Lambda:~\mathcal{B}(H)\rightarrow
\mathcal{B}(H)$ that
admits a form of Kraus operator representation, i.e.,
\begin{eqnarray}
\Lambda(\cdot)=\sum\limits_{i}X_i(\cdot) X_i^\dag\label{b}
\end{eqnarray}
where $X_i$'s are operators acting on $H$  with
$\sum\limits_{i}X_i^\dag X_i=I$.

In particular, a channel
$\Lambda$ acting on an $n-$dimensonal quantum system is called an
\emph{isotropic channel} if it has the form
\begin{eqnarray}
\Lambda(\cdot)=t\Gamma(\cdot)+(1-t){\rm Tr}(\cdot)\frac{I}{n},\label{bb}
\end{eqnarray}
where $\Gamma$ is either a unitary operation or unitarily equivalent
to transpose (also see in Ref.~\cite{Hu}). Parameter $t$ is chosen to make sure that $\Lambda$ is
a trace-preserving completely positive linear map. If $t$ in
Eq.(\ref{bb}) is nonzero, we call $\Lambda$ a nontrivial isotropic
channel.
It is known by \cite{Hu}, $\frac{-1}{n-1}\leq t\leq 1$
when $\Gamma$ is a unitary operation, and $\frac{-1}{n-1}\leq t\leq
\frac{1}{n+1}$ when $\Gamma$ is unitarily equivalent to transpose.
If $t=0$, $\Lambda$ is the \emph{completely depolarizing channel}, namely, $\Lambda(\mathcal{S}(H))=\{\frac{1}{n}I\}$.
A channel $\Lambda$ is called a \emph{completely decohering
channel} (or semi-classical channel) if $\Lambda(\mathcal{B}(H))$ is
commutative. In general, the completely depolarizing channel is viewed as
a special case of completely decohering one.

\if
or equivalently, $\Lambda$ nullifies QD in any states.
\fi

Quantum discord, as a
quantum correlation of bipartite system, is initially introduced by Ollivier and Zurek
\cite{Ollivier} and by Henderson and Vedral \cite{Henderson}.
We denote by A+B the bipartite system shared by Alice and Bob.
Let $H_A$ and $H_B$ be the complex Hilbert spaces that describing the subsystem of Alice and Bob, respectively.
Then $H_A\otimes H_B$ corresponds to the composite system A+B.
Recall that, for a state $\rho\in{\mathcal S}(H_A\otimes
H_B)$, the quantum discord of $\rho$ (up to part B) is defined by
\begin{eqnarray}
D_B(\rho):=\min_{\Pi^b}\{I(\rho)-I(\rho|\Pi^b)\},
\end{eqnarray}
where, the minimum is taken over all local von Neumann measurements
$\Pi^b$,
$I(\rho):=S(\rho_A)+S(\rho_B)-S(\rho)$
is interpreted as the quantum mutual information,
$S(\rho):=-{\rm Tr}(\rho\log\rho)$
is the von Neumann entropy,
$I(\rho|\Pi^b)\}:=S(\rho_A)-S(\rho|\Pi^b)$,
$S(\rho|\Pi^b):=\sum_kp_kS(\rho_k)$,
and
$\rho_k=\frac{1}{p_k}( I_A\otimes\Pi_k^b)\rho(I_A\otimes\Pi_k^b )$
with $p_k={\rm Tr}[(I_A\otimes\Pi_k^b)\rho(I_A\otimes\Pi_k^b)]$,
$k=1$, 2, $\dots$, $\dim H_B$.
\if
It is easy to check that a channel $\Lambda$ is completely decohering
if and only if $I_A\otimes \Lambda$ nullifies QD in any,
where $I_A$ denotes the identity map acting on part A.
The geometric measure of
quantum discord (GMQD) of a state (up to part B) is originally
introduced in Ref.~\cite{Dakic} as
\begin{eqnarray*}
D_B^G(\rho):=\min_{\chi}\|\rho-\chi\|_2^2
\end{eqnarray*}
with $\chi$ runs over all zero QD states , where $\|\cdot\|_2$
stands for the Hilbert-Schmidt norm (that is $\|A\|_2=[{\rm
Tr}(A^\dag A)]^{\frac{1}{2}}$).  Symmetrically, one can define QD and
GMQD of a state up to part A.\fi

Let $\phi:\mathcal{B}(H)\rightarrow\mathcal{B}(H)$ be a map.
Throughout this paper, we say that (i) $\phi$ preserves normality if
$\phi$ maps normal operators to normal operators, namely,
$A\in{\mathcal B}(H)$ is normal implies that $\phi(A)$ is normal
(here, an operator $A\in{\mathcal B}(H)$ is
called a normal operator if $AA^\dag=A^\dag A$);
(ii) $\phi$ preserves normality in both directions if $A\in{\mathcal
B}(H)$ is normal if and only if $\phi(A)$ is normal; (iii) $\phi$
preserves commutativity (or $\phi$ is a commutativity-preserving map)
if $[A,B]=AB-BA=0$ implies
$[\phi(A),\phi(B)]=0$ for $A$, $B\in\mathcal{B}(H)$; (iv) $\phi$
preserves commutativity in both directions if
$[A,B]=0\Leftrightarrow[\phi(A),\phi(B)]=0$ for $A$,
$B\in\mathcal{B}(H)$; (v) $\phi$ preserves commutativity for
hermitian operators (resp. quantum states) if $[A,B]=0$ implies $[\phi(A),\phi(B)]=0$ for
hermitian operators (resp. quantum states) $A$ and $B$ in $\mathcal{B}(H)$ (resp. $\mathcal{S}(H)$); (vi) $\phi$
preserves commutativity in both directions for hermitian operators (resp. quantum states)
if $[A,B]=0\Leftrightarrow[\phi(A),\phi(B)]=0$ holds for hermitian
operators (resp. quantum states) $A$ and $B$  in $\mathcal{B}(H)$ (resp. $\mathcal{S}(H)$). We say that a local
channel cannot create QD for zero QD states
if $D_B(\rho)=0\Rightarrow D_B((I_A\otimes \Lambda)\rho)=0$, where $I_A$ denotes the
identity map acting on part A. For
$A\in\mathcal{B}(H)$, $A^T$ denotes the transpose of $A$ relative to
an arbitrarily fixed basis.

\section{Local channels that cannot create QD for zero QD states}

This section is devoted to discussing the local channels that cannot
create QD for zero QD states. We first
consider the qudit case with $d\geq3$ and then
discuss the qubit case.

\subsection{The qudit case ($d\geq3$)}

The following is the main result of this subsection.

{\it Theorem 1}.\quad Let $H_A$ and $H_B$ be complex Hilbert spaces
with $\dim H_A=m\geq 2$ and $\dim H_B=n\geq3$, and let $\Lambda$ be
a channel acting on subsystem B. Then the following statements
are equivalent.

(1) $D_B(\rho)=0 \Rightarrow D_B((I_A\otimes\Lambda)\rho)=0$.

\if
(1$^\prime$) $D_B^G(\rho)=0 \Rightarrow
D_B^G((I_A\otimes\Lambda)\rho)=0$.
\fi

(2) $\Lambda$ preserves commutativity for hermitian operators.

(2$^\prime$) $\Lambda$ preserves commutativity.

(2$^{\prime\prime}$) $\Lambda$ preserves normality.

(2$^{\prime\prime\prime}$) $\Lambda$ preserves commutativity for quantum states.

(3) Either (a) $\Lambda$ is a completely decohering channel
or (b) $\Lambda$ is a nontrivial isotropic channel.

\if there exists a unitary operator $U$ and a real number $t\neq 0$
such that $\Lambda$ has one of the following forms:

~~(i) $\Lambda(A)=tU A U^\dag+\frac{1-t}{n}{\rm Tr}(A)I_B$ for
all $A$ in $\mathcal{B}(H_B)$, where $\frac{-1}{n-1}\leq t \leq 1$.

~~(ii) $\Lambda(A)=tU A^TU^\dag +\frac{1-t}{n}{\rm Tr}(A)I_B$
for all $A$ in $\mathcal{B}(H_B)$, where $\frac{-1}{n-1}\leq t \leq \frac{1}{n+1}$.\fi

It is easily checked that the maps of the form  (a) in item (3) are
channels. The equivalence of (1) and (3) implies that a local
channel cannot create QD for zero QD states if and only if either
it is a completely decohering channel, or it is an isotropic
channel. Therefore, our Theorem 1 particularly solves
affirmatively the conjecture proposed in Ref.~\cite{Hu}.

{\it Proof of Theorem 1}.\quad(3)$\Rightarrow$(2) and (2)$\Rightarrow$(2$^{\prime\prime\prime}$) are obvious.
(2$^{\prime\prime\prime}$)$\Rightarrow$(2) holds since two hermitian matrices $A$, $B$ are commutative if and only if
$[A^+,B^+]=[A^+,B^-]=[B^+,A^+]=[B^+,A^-]=0$, where $A^+$ and $A^-$ are the positive part and the negative part of
$A$ respectively, $B^+$ and $B^-$ are the positive part and the negative part of $B$ respectively (Note that in such a decomposition, $A^{+,-}\geq0$, $B^{+,-}\geq0$ and $[A^+,A^-]=[B^+,B^-]=0$).
\if
Since $D_B(\rho)=0$ if and only if $D_B^G(\rho)=0$ \cite{GuoHou},
we have (1)$\Leftrightarrow$(1$^\prime$) directly.
\fi

By \cite[Corollary 1]{Choi87} we know that if
$\phi:\mathcal{B}(H_B)\rightarrow\mathcal{B}(H_B)$ is a
hermitian-preserving linear map (namely, $\phi(A^\dag)=\phi(A)^\dag$
for every $A$), then $\phi$ preserves commutativity if and only if
it preserves normality, and in turn, if and only if it preserves
commutativity for hermitian operators. Hence,
(2)$\Leftrightarrow$(2$^\prime$)$\Leftrightarrow$(2$^{\prime\prime}$)
is immediate since the channels are hermitian-preserving.

(2)$\Rightarrow$(3). Denote by $\mathcal{H}_n$ the real linear
space of all $n\times n$ hermitian complex matrices.
Let $\Lambda$
as in Eq.~(\ref{b}).
Then
$\Lambda$ is a commutativity-preserving map on $\mathcal{H}_n$.
Let $\mathcal{M}_n$ be the algebra of all $n$ by $n$ matrices.
 By \cite[Theorem 3]{Choi87},
if $\phi$ is a hermitian-preserving (i.e.,
$\phi(A^\dag)=\phi(A)^\dag$) linear map on ${\mathcal M}_n$ which
also preserves commutativity for hermitian matrices, then either
$\phi(\mathcal{M}_n)$ is commutative, or there exist a unitary
matrix $U$, a linear functional $f$ on $\mathcal{M}_n$, and a real number
$t$ such that $\phi$ has one of the following forms: (i)
$\phi(A)=tUAU^\dag+f(A)I$ for all $A$ in $\mathcal{M}_n$; (ii)
$\phi(A)=tUA^TU^\dag +f(A)I$ for all $A$ in $\mathcal{M}_n$. Note
that $\Lambda$ is a hermitian-preserving linear map. Therefore,
either $\Lambda(\mathcal{B}(H_B))$ is commutative, or there exists a
unitary operator $U\in{\mathcal B}(H_B)$ and a real number $t$ such that
$\Lambda$ has one of the following forms: (i) $\Lambda(A)=tU A
U^\dag+f(A)I_B$ for all $A$ in $\mathcal{B}(H_B)$. (ii)
$\Lambda(A)=tU A^TU^\dag +f(A)I_B$ for all $A$ in
$\mathcal{B}(H_B)$. Note that $\mathcal{B} (H_B)$ can be regarded as
a Hilbert space with the Hilbert-Schmidt inner product
\begin{eqnarray*}
\langle X|Y\rangle:={\rm Tr}(X^\dag Y).
\end{eqnarray*}
It turns out that there exists an operator $W\in{\mathcal B}(H_B)$
such that $f(A)={\rm Tr}(WA)$ holds for all $A\in{\mathcal B}(H_B)$.
Considering the action of $\Lambda$ on $\mathcal{S}(H_B)$ one   gets
that $t+nf(\rho)=1$ for all $\rho\in\mathcal{S}(H_B)$. It follows that
${\rm Tr}(W\rho)\equiv\frac{1-t}{n}$
for all $\rho$ in $\mathcal{S}(H_B)$.
 \if false Write $W=[w_{ij}]$.
It is easy to compute that $w_{ii}=\frac{1-t}{n}$. We show $W$ is a
hermitian matrix. Or else, we assume with no loss of generality that
$w_{i_0j_0}\neq \bar{w}_{j_0i_0}$ for some $i_0$ and $j_0$. Let
$\alpha$ and $\beta$ are complex numbers with
$\bar{\alpha}\beta(w_{i_0j_0}-\bar{w}_{j_0i_0})$ is not a real
number that satisfies
$\bar{\alpha}\beta\neq\bar{w}_{i_0j_0}-{w}_{j_0i_0}$. Let
$A_0=[a_{ij}]$ with $a_{i_0i_0}=|\alpha|^2$,
$a_{i_0j_0}=\alpha\bar{\beta}$,
 $a_{j_0i_0}=\bar{\alpha}\beta$, $a_{j_0j_0}=|\beta|^2$
 and $a_{ij}=0$ for all $i\neq i_0$, $j\neq i_0$
$i\neq j_0$, $j\neq j_0$. Direct calculation shows that ${\rm
Tr}(WA_0)=\frac{1-t}{n}|\alpha|^2+\bar{\alpha}\beta
w_{i_0j_0}+\alpha\bar{\beta} w_{j_0i_0}+\frac{1-t}{n}|\beta|^2$.
Note that ${\rm Tr}(WA_0)\in\mathbb{R}$ since $A_0$ is hermitian and
$\Lambda$ maps hermitian matrix to hermitian matrix. That is
$\frac{1-t}{n}|\alpha|^2+\bar{\alpha}\beta
w_{i_0j_0}+\alpha\bar{\beta} w_{j_0i_0}+\frac{1-t}{n}|\beta|^2$ is a
real number. Hence $\bar{\alpha}\beta w_{i_0j_0}+\alpha\bar{\beta}
w_{j_0i_0} =(\bar{\alpha}\beta w_{i_0j_0}-\bar{\alpha}\beta
\bar{w}_{j_0i_0}) +(\bar{\alpha}\beta
\bar{w}_{j_0i_0}+\alpha\bar{\beta} w_{j_0i_0})$ is a real number,
which implies that $\bar{\alpha}\beta w_{i_0j_0}-\bar{\alpha}\beta
\bar{w}_{j_0i_0}=\bar{\alpha}\beta(w_{i_0j_0}-\bar{w}_{j_0i_0})$ is
a real number, a contradiction. Thus $W$ is hermitian. It follows
that\fi Consequently, ${\rm Tr}(WA)=\frac{1-t}{n}{\rm Tr}(A)$ holds
for all $A\in{\mathcal B}(H_B)$, which implies that
\begin{eqnarray*}
W\equiv\frac{1-t}{n}I_B.
\end{eqnarray*}
From Ref.~\cite{Hu}, $t$
satisfies $\frac{-1}{n-1}\leq t \leq 1$ when $\Lambda(A)=tU A
U^\dag+\frac{1-t}{n}{\rm Tr}(A)I_B$ and $\frac{-1}{n-1}\leq t \leq
\frac{1}{n+1}$ when $\Lambda(A)=tU A^T U^\dag+\frac{1-t}{n}{\rm
Tr}(A)I_B$, which guarantees that $\Lambda$ is completely positive.
That is, $\Lambda$ is an isotropic channel.
\if Thus $\Lambda$ has the form stated in (b) of item (3) whenever
$t\neq 0$.\fi
If $t\neq0$, it is the nontrivial isotropic channel, i.e., the case of item (b);
If $t=0$, it is obvious that $\Lambda$ is a complete depolarizing channel, i.e., a special case of item (a).

If $\Lambda(\mathcal{B}(H_B))$ is commutative, then $\Lambda$
is a completely decohering channel.
That is, (a) of item (3) holds.

\if It is worth mentioning here that the maps   stated in (a) of
item (3) are channels.  In order to see it, we let $\{|k'\rangle\}$
be the canonical computational basis of $\mathbb{C}^n\cong H_B$ and
let $E_{kl}=|k'\rangle\langle l'|$. Let $V$ be a unitary matrix that
transforms $|k'\rangle$ to $|e_k\rangle$. Write
\begin{eqnarray*}
U_iW_iU_i^\dag=D_i={\rm diag}(d_1^{(i)},d_2^{(i)},\dots,d_n^{(i)}),
\end{eqnarray*}
where $U_i$s are $n$ by $n$ unitary matrices.
One can easily check that
\begin{eqnarray*}
\begin{array}{rl}
\Lambda(A)
=&\sum\limits_i|e_i\rangle\langle e_i|V(E_{i1}+E_{i2}+\dots+E_{in})\\
&\cdot(\sum\limits_jd_j^{(i)}E_{jj}U_iAU_i^\dag E_{jj})\\
&\cdot(E_{1i}+E_{2i}+\dots+E_{ni})V^\dag|e_i\rangle\langle e_i|,
\end{array}
\end{eqnarray*}
which implies that $\Lambda$ is completely positive according to the
condition of completely positive linear map proposed in
\cite{Choi75}. (Remark here that $\sum_iX_iX_i^\dag$ is a diagonal
matrix with respect to the basis $\{|e_i\rangle\}$.)\fi

 (1)$\Rightarrow$(2).
Let $\{|i\rangle\}$ be an orthonormal basis of $H_A$. Then any state
$\rho$ acting on $H_A\otimes H_B$ can be represented by
\begin{eqnarray}
\rho=\sum_{i,j}E_{ij}\otimes B_{ij},\label{l}
\end{eqnarray}
where $E_{ij}=|i\rangle\langle j|$ and $B_{ij}$'s are are operators acting on $H_B$, and
\begin{eqnarray*}
(I_A\otimes\Lambda)\rho=\sum_{i,j}E_{ij}\otimes \Lambda(B_{ij}).
\end{eqnarray*}
We proved in Ref.~\cite{GuoHou} that $D_B(\rho)=0$ if and only if
$B_{ij}$'s are mutually commuting normal operators.
It follows from $D_B(\rho)=0\Rightarrow D_B((I_A\otimes \Lambda)\rho)=0$
that $\Lambda$ preserves normality, and thus, preserves
commutativity for hermitian operators according to \cite[Corollary 1]{Choi87}.
In fact, for any normal operator $A\in\mathcal{B}(H_B)$, there exist
positive operators $C,D\in{\mathcal B}(H_B)$ such that $A,C,D$ are
mutually commutating and
\begin{eqnarray*}
\begin{array}{rcl}
\rho_0
&=&\frac{1}{{\rm Tr}(C+D)}(E_{11}\otimes C+E_{12}\otimes A\\
&&
+E_{21}\otimes A^\dag+E_{22}\otimes D)
\end{array}
\end{eqnarray*}
is a state. Moreover, by the result in Ref.~\cite{GuoHou} mentioned above,
we have $D_B(\rho_0)=0$. Thus  $D_B((I_A\otimes\Lambda)\rho_0)=0$,
which implies that $\Lambda(A), \Lambda (C), \Lambda (D), \Lambda
(A^\dag)=\Lambda (A)^\dag$ are mutually commuting normal operators.
In particular, $\Lambda (A)$ is normal.

(2)$\Rightarrow$(1). Since $\Lambda$ is a hermitian-preserving
linear map, (1) implies that $\Lambda$ preserves normality and
commutativity. Therefore, if $B_{ij}$'s are mutually commuting normal
operators, then $\Lambda(B_{ij})$'s also are mutually commuting normal
operators. Now, by Ref.~\cite{GuoHou}, it is obvious that $D_B(\rho)=0
\Rightarrow D_B((I_A\otimes\Lambda)\rho)=0$. \hfill$\blacksquare$

Furthermore, we have

{\it Proposition 1}.\quad Let $H_A$ and $H_B$ be complex Hilbert
spaces with $\dim H_A=m\geq2$ and $\dim H_B=n\geq3$, and let $\Lambda$
be a channel acting on subsystem B. Then the following
statements are equivalent.

(1) $D_B(\rho)=0\Leftrightarrow D_B((I_A\otimes\Lambda)\rho)=0$.

\if (1$^\prime$) $D_B^G(\rho)=0\Leftrightarrow
D_B^G((I_A\otimes\Lambda)\rho)=0$.\fi

(2) $\Lambda$ preserves commutativity in both directions for
hermitian operators.

(2$^\prime$) $\Lambda$ preserves commutativity in both directions.

(2$^{\prime\prime}$) $\Lambda$ preserves normality in both
directions.

(2$^{\prime\prime\prime}$) $\Lambda$ preserves commutativity in both direction for quantum states.

(3) $\Lambda$ is a nontrivial isotropic channel. \if There exists a unitary matrix $U$ and a nonzero number $t$
such that $\Lambda$ has one of the following forms:

~~(i) $\Lambda(A)=tU A U^\dag+\frac{1-t}{n}{\rm Tr}(A)I_B$ for all
$A$ in $\mathcal{B}(H_B)$, where $\frac{-1}{n-1}\leq t \leq 1$.

~~(ii) $\Lambda(A)=tU A^TU^\dag +\frac{1-t}{n}{\rm Tr}(A)I_B$ for
all $A$ in $\mathcal{B}(H_B)$, where $\frac{-1}{n-1}\leq t \leq
\frac{1}{n+1}$.\fi

{\it Proof}.\quad (2)$\Leftrightarrow$(2$^\prime$), (2)$\Leftrightarrow$(2$^{\prime\prime\prime}$),
(3)$\Rightarrow$(1), (3)$\Rightarrow$(2) and
(3)$\Rightarrow$(2$^\prime$) are obvious.

(2)$\Leftrightarrow$(2$^{\prime\prime}$) is easily checked by the
fact that $A$ is normal if and only if it can be written as
$A=X+iY$ with $X$ and $Y$  hermitian and $[X,Y]=0$.

(1)$\Rightarrow$(3). According to (1)$\Leftrightarrow$(3) of
Theorem 1, $\Lambda$ admits the form of item (3) in Theorem 1.
It is
clear that the case (a) of item (3) cannot occur since
the completely decohering channel nullifies QD in any state~\cite{Yusixia}. So, $\Lambda$ is a nontrivial isotropic channel.

(2)$\Rightarrow$(3). According to the proof of
(2)$\Rightarrow$(3) in Theorem 1, we know that
 $\Lambda$ admits the form as in item (3) of
 Theorem 1. It is immediate that the case of `$\Lambda(\mathcal{B}(H_B))$ is
 commutative' cannot occur.\hfill$\blacksquare$

\if
Proposition 1 implies that, for the case $n>2$, a local channel
$\Lambda$ neither creates nor vanishes the quantum correlations
measured by QD for zero QD states if and only if it preserves
commutativity in both directions, and  in turn, if and only if it is
a nontrivial isotropic channel.

 Then what happens for the qubit
case?
\fi

\subsection{The qubit case}

We now turn to the discussion of the qubit case, that is, $\dim H_B=2$.
We will show that the form of commutativity-preserving unital channel for
qubit system is different from the higher dimensional case.  Thus,
the local channel of qubit system that preserve zero QD states has
 different forms from that of higher dimensional systems.

\if Let $U\in{\mathcal B}(H_B)$ be   unitary and $-1\leq t\leq1$. Let
$\Lambda :{\mathcal B}(H_B)\rightarrow{\mathcal B}(H_B)$ be a
channel. It is clear that, if $\Lambda(A)=tU A
U^\dag+\frac{1-t}{2}{\rm Tr}(A)I_B$ for all $A$ in
$\mathcal{B}(H_B)$ or $\Lambda(A)=tU A^TU^\dag +\frac{1-t}{2}{\rm
Tr}(A)I_B$ for all $A$ in $\mathcal{B}(H_B)$ when $-1\leq
t\leq\frac{1}{3}$, then  $D_B((I_A\otimes\Lambda)\rho)=0$ whenever
$D_B(\rho)=0$; if $t\neq0$ and $\Lambda(A)=tU A
U^\dag+\frac{1-t}{2}{\rm Tr}(A)I_B$ for all $A$ in
$\mathcal{B}(H_B)$ or $\Lambda(A)=tU A^TU^\dag +\frac{1-t}{2}{\rm
Tr}(A)I_B$ for all $A$ in $\mathcal{B}(H_B)$ when $-1\leq
t\leq\frac{1}{3}$, then $D_B(\rho)=0$ if and only if
$D_B((I_A\otimes\Lambda)\rho)=0$.\fi

Compared with Theorem 1, the main result of this
subsection is the following.

{\it Theorem 2}.\quad Let $H_A$ and $H_B$ be complex Hilbert spaces
with $\dim H_A=m\geq2$ and $\dim H_B=2$, and let $\Lambda$ be a
channel acting on subsystem B. Then the following statements are
equivalent.

(1) $D_B(\rho)=0 \Rightarrow D_B((I_A\otimes\Lambda)\rho)=0$.

\if
(1$^\prime$) $D_B^G(\rho)=0 \Rightarrow
D_B^G((I_A\otimes\Lambda)\rho)=0$.
\fi

(2) $\Lambda$ preserves commutativity for hermitian operators.

(2$^\prime$) $\Lambda$ preserves commutativity.

(2$^{\prime\prime}$) $\Lambda$ preserves normality.

(2$^{\prime\prime\prime}$) $\Lambda$ preserves commutativity for quantum states.

(3) Either (a) $\Lambda$ is a completely decohering channel; or
(b) for  any orthonormal basis $\{|e_1\rangle,|e_2\rangle\}$
of $H_B$, there exist a unitary operator $U\in{\mathcal B}(H_B)$,
real numbers $0\leq\lambda\leq1$, and complex numbers
$\alpha,\beta,\gamma$ so that  $\left(\begin{array}{cc} \alpha &\beta \\
\gamma& -\alpha \end{array}\right)$ is
contractive,  such that, with respect to the space decomposition
$H_B={\mathbb C}|e_1\rangle\oplus {\mathbb C}|e_2\rangle$,
\begin{widetext}
\begin{eqnarray}
\Lambda\left(\left(\begin{array}{cc}
a_{11}&a_{12}\\
a_{21}&a_{22}
\end{array}
\right)\right)
= U\left[ \left(\begin{array}{cc}
\lambda a_{11}+(1-\lambda)a_{22}&0\\
0&(1-\lambda)a_{11}+\lambda a_{22}
\end{array}
\right)+a_{12}X+a_{21}X^\dag\right]U^\dag,\label{8}
\end{eqnarray}
\end{widetext}
for all $A=\left(\begin{array}{cc}
a_{11}&a_{12}\\
a_{21}&a_{22}
\end{array}
\right)$, where
\begin{eqnarray*}
X=
\left(\begin{array}{cc}
\sqrt{\lambda(1-\lambda)} \alpha&\lambda\beta\\
(1-\lambda)\gamma&-\sqrt{\lambda(1-\lambda)}\alpha
\end{array}
\right)
\end{eqnarray*}
with $|\beta|+|\gamma|\neq0$, $\beta\neq0$ when $\lambda=1$ and
$\gamma\neq0$ when $\lambda=0$.

Theorem 2 depicts the commutativity-preserving unital qubit channel in detail, which improves
the Theorem 1 in ~\cite{Streltsov} proposed by Streltsov {\it et al.}

{\it Proof of Theorem 2}.\quad We only need to check
the implication (2)$\Rightarrow$(3). By
 Theorem 5 (see in Appendix), and noting that $\Lambda$ is completely
positive and trace-preserving, we can know that $\Lambda$ is a completely decohering channel
if the range of
$\Lambda$ is commutative.

Assume that the range of $\Lambda$ is not commutative. Take an
orthonormal basis $\{|e_1\rangle, |e_2\rangle\}$ of $H_B$ and denote
$E_{ij}=|e_i\rangle\langle e_j|$, $i,j=1,2$. Then Theorem 5 ensures
that there exist a unitary operator $U$ on $H_B$ and nonnegative
real numbers $\lambda_1, \lambda_2, \mu_1,\mu_2$ with
$\lambda_1+\mu_1=\lambda_2+\mu_2$ such that, with respect to the space decomposition
$H_B=\mathbb{C}|e_1\rangle+\mathbb{C}|e_2\rangle$,
\begin{eqnarray*} \Lambda(E_{11})= U\left(\begin{array}{cc}
\lambda_1&0\\
0& \lambda_2
\end{array}
\right)U^\dag,
\Lambda(E_{22})= U\left(\begin{array}{cc}
\mu_1 &0\\
0&\mu_2
\end{array}
\right)U^\dag.
\end{eqnarray*}
As ${\rm Tr}(\phi(E_{11}))={\rm Tr}(\phi(E_{22}))=1$,
 we must have
$\lambda_1=\mu_2=1-\lambda_2=1-\mu_1\geq 0$ and
$\lambda_1+\mu_1=1$. Let $\lambda_1=\lambda$. Then
$0\leq\lambda\leq1$ and
\begin{eqnarray} &\Lambda(E_{11})= U\left(\begin{array}{cc}
\lambda&0\\
0&1-\lambda
\end{array}
\right)U^\dag,& \label{12} \\
&\Lambda(E_{22})= U\left(\begin{array}{cc}
1-\lambda&0\\
0&\lambda
\end{array}
\right)U^\dag.& \label{13}
\end{eqnarray}

Note that ${\rm Tr}(\Lambda(E_{12}))=0$. So, there are complex
numbers
$x,y,z$ such that $\Lambda (E_{12})=U\left(\begin{array}{cc} x& y\\
z & -x\end{array}\right)U^\dag$ and
$\Lambda(E_{21})=\Lambda(E_{12})^\dag$.

It is well-known by a theorem of Choi~\cite{Choi75} that the map $\Lambda$ is completely
positive if and only if the block matrix $[\Lambda(E_{ij})]$ is
positive. It follows from Eqs.~(\ref{12}) and (\ref{13}) that
\begin{widetext}
\begin{eqnarray*}
\left(\begin{array}{cc}\Lambda(E_{11})& \Lambda(E_{12})\\
\Lambda(E_{21})&\Lambda(E_{22})\end{array}\right)
=\left(\begin{array}{cc}
U&0\\
0&U
\end{array}
\right)
\left(\begin{array}{cc|cc}
\lambda&0&x&y\\
0&1-\lambda&z&-x\\
\hline
\bar{x}&\bar{z}&1-\lambda&0\\
 \bar{y} &-\bar{x}&0&\lambda
\end{array}\right)\left(\begin{array}{cc}
U^\dag&0\\
0&U^\dag
\end{array}
\right).
\end{eqnarray*}
\end{widetext}
Let
$J=\left(\begin{array}{cc|cc}
\lambda&0&x&y\\
0&1-\lambda&z&-x\\
\hline
\bar{x}&\bar{z}&1-\lambda&0\\
 \bar{y} &-\bar{x}&0&\lambda
\end{array}\right)$.
Then $\Lambda$ is completely positive if and only if
 $J\geq 0$, and in turn, from Theorem 1.1 in Ref.~\cite{Hougao}, if
and only if there exists a contractive matrix
$S=\left(\begin{array}{cc} \alpha &\beta \\ \gamma &\eta
\end{array}\right)$ such that
\begin{widetext}
\begin{eqnarray*}
\left(\begin{array}{cc} x & y \\ z &-x \end{array}\right)
=\left(\begin{array}{cc} \sqrt{\lambda} & 0 \\ 0
&\sqrt{1-\lambda}
\end{array}\right)\left(\begin{array}{cc} \alpha & \beta \\ \gamma &\eta \end{array}\right)\left(\begin{array}{cc} \sqrt{1-\lambda} & 0 \\ 0 &\sqrt{\lambda}
\end{array}\right)
= \left(\begin{array}{cc} \sqrt{\lambda (1-\lambda)}\alpha & {\lambda}\beta \\
 (1-\lambda)\gamma &\sqrt{\lambda (1-\lambda)}\eta
\end{array}\right).
\end{eqnarray*}
\end{widetext}
It follows that $\eta=-\alpha$ and
\begin{eqnarray*}
X=U^\dag\Lambda(E_{12})U
=\left(\begin{array}{cc} \sqrt{\lambda (1-\lambda)}\alpha & {\lambda}\beta \\
 (1-\lambda)\gamma &-\sqrt{\lambda (1-\lambda)}\alpha
\end{array}\right).
\end{eqnarray*}

If $|\beta|+|\gamma|=0$, or $\beta=0$ when $\lambda=1$, or
$\gamma=0$ when $\lambda=0$, then the channel reduces to a completely decohering one.
Now it is
clear that $\Lambda$ has the form of Eq.~(\ref{8}). This completes the
proof. \hfill$\blacksquare$

It is worth highlighting that (i) a channel with the form as in Eq.~(\ref{8}) does not preserves
commutativity in both directions necessarily, which is different from the higher dimensional case;
(ii) the isotropic qubit channel is only a special case of
(b) in item (3) of Theorem 2. The map $\Lambda$ has the form
$\Lambda(A)=tUAU^\dag+\frac{1-t}{2}{\rm Tr}(A)I$ for all $A$ if and
only if  $t=2\lambda-1$,
$\beta=\frac{2\lambda-1}{\lambda}$ and
$\alpha=\gamma=0$; $\Lambda$ has the form
$\Lambda(A)=tUA^TU^\dag+\frac{1-t}{2}{\rm Tr}(A)I$ for all $A$ if
and only if  $t=2\lambda-1$,
$\gamma=\frac{2\lambda-1}{1-\lambda}$ and
$\alpha=\beta=0$. So there are many commutativity-preserving unital
channels for qubit system that are neither isotropic nor
completely decohering. This is quite different from the case
of $n\geq3$ as stated in Theorem 1.

Going further, we have

{\it Proposition 2}.\quad Let $H_A$ and $H_B$ be complex Hilbert
spaces with $\dim H_A=m$ and $\dim H_B=2$, and let $\Lambda$ be a
channel acting on subsystem B. Then the following statements
are equivalent.

(1) $D_B(\rho)=0\Leftrightarrow D_B((I_A\otimes\Lambda)\rho)=0$.

\if
(1$^\prime$) $D_B^G(\rho)=0\Leftrightarrow
D_B^G((I_A\otimes\Lambda)\rho)=0$.
\fi

(2) $\Lambda$ preserves commutativity in both directions for
hermitian operators.

(2$^\prime$) $\Lambda$ preserves commutativity in both directions.

(2$^{\prime\prime}$) $\Lambda$ preserves normality in both
directions.

(2$^{\prime\prime\prime}$) $\Lambda$ preserves commutativity in both directions for quantum states.

(3) \if Taking  any orthonormal basis $\{|e_1\rangle,|e_2\rangle\}$ of
$H_B$, there exists a unitary operator $U\in{\mathcal B}(H_B)$, real
numbers $\lambda$ with $0\leq\lambda\leq1$,
and complex numbers
$\alpha,\beta,\gamma$ so that  $\left(\begin{array}{cc} \alpha &\beta \\
\gamma& -\alpha \end{array}\right)$ is
contractive,  such that, with respect to the space decomposition
$H_B={\mathbb C}|e_1\rangle\oplus {\mathbb C}|e_2\rangle$,
\begin{eqnarray}\begin{array}{rl}
& \Lambda(\left(\begin{array}{cc}
a_{11}&a_{12}\\
a_{21}&a_{22}
\end{array}
\right))\\
= &U[ \left(\begin{array}{cc}
\lambda a_{11}+(1-\lambda)a_{22}&0\\
0&(1-\lambda)a_{11}+\lambda a_{22}
\end{array}
\right) \\ & +a_{12}X+a_{21}X^\dag]U^\dag,\end{array}
\end{eqnarray}
for all $A=\left(\begin{array}{cc}
a_{11}&a_{12}\\
a_{21}&a_{22}
\end{array}
\right)$, where
\begin{eqnarray*}
X=
\left(\begin{array}{cc}
\sqrt{\lambda(1-\lambda)}\alpha&\lambda\beta\\
(1-\lambda)\gamma&-\sqrt{\lambda(1-\lambda)}\alpha
\end{array}
\right)
\end{eqnarray*}
with
\fi
$\Lambda$ has the form as Eq.~{(\ref{8})} with $X$ satisfying
 $\lambda\neq \frac{1}{2}$, $|\beta|+|\gamma|\neq0$;
$\gamma\neq0$ when $\lambda=0$;
$\beta\neq0$ when $\lambda=1$;
$\lambda|\beta|\neq(1-\lambda)|\gamma|$ when $\alpha=0$,
$\beta\gamma\neq0$ and $\lambda(1-\lambda)\neq0$; and
$\lambda|\beta|\neq(1-\lambda)|\gamma|$ or
$\lambda\bar{\beta}\alpha\neq(1-\lambda)\gamma\bar{\alpha}$
when $\alpha\beta\gamma\neq0$ and $\lambda(1-\lambda)\neq0$.
\if $\lambda\beta-(1-\lambda)\bar{\gamma}\neq0$ when
$\lambda\beta\pm(1-\lambda)\gamma$ is a pure imaginary and
$\lambda\beta+(1-\lambda)\bar{\gamma}\neq0$ when
$\lambda\beta\pm(1-\lambda)\gamma$ is a real number.\fi

{\it Proof}.\quad We only need to check that (2)$\Leftrightarrow$(3).

(2)$\Rightarrow$(3). As $\Lambda$ preserves commutativity in both
directions, $\Lambda$ has the form as Eq.~{(\ref{8})} and is
injective. The injectivity reveals that $X$ satisfies the conditions as in term (3) above.
\if
$\lambda\neq \frac{1}{2}$, $|\beta|+|\gamma|\neq0$; $\gamma\neq0$ when $\lambda=0$;
$\beta\neq0$ when $\lambda=1$;
$\lambda|\beta|\neq(1-\lambda)|\gamma|$
when $\alpha=0$, $\beta\gamma\neq0$ and $\lambda(1-\lambda)\neq0$; and
$\lambda|\beta|\neq(1-\lambda)|\gamma|$ or
$\lambda\bar{\beta}\alpha\neq(1-\lambda)\gamma\bar{\alpha}$
when $\alpha\beta\gamma\neq0$ and $\lambda(1-\lambda)\neq0$.
\fi
 That is,
(3) holds.

(3)$\Rightarrow$(2). By Theorem 2, (3) implies that $\Lambda$
preserves commutativity. Moreover, the conditions ensure that
$\Lambda$ is injective. For any $A\in{\mathcal B}(H_B)$, denote by
$\{A\}^\prime$ the commutant of $A$, that is,
$\{A\}^\prime=\{B\in{\mathcal B}(H_B) : AB=BA\}$. Then $\dim
\{\Lambda(A)\}^\prime=\dim\{A\}^\prime$. This entails that
$[\Lambda(A),\Lambda(B)]=0\Rightarrow [A,B]=0$. So, $\Lambda$
preserves commutativity in both directions.\hfill$\blacksquare$

Proposition 1 and Proposition 2 imply that a local channel neither creates nor vanishes
the quantum correlation measured by quantum discord if and only if it preserves commutativity in
both directions, and in turn, if and only if it is one-to-one and it outputs commutative
states whenever the input states are commutative.

\begin{figure}
    \centering
    \includegraphics[width=0.35\textwidth]{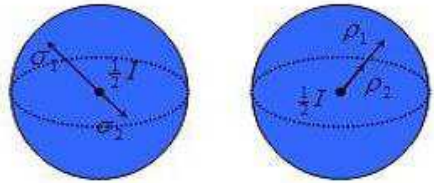}
    \caption{$\rho_1$ and $\rho_2$ are commutative.}
    \label{fig: 1}
  \end{figure}

   \begin{figure}
    \centering
    \includegraphics[width=0.35\textwidth]{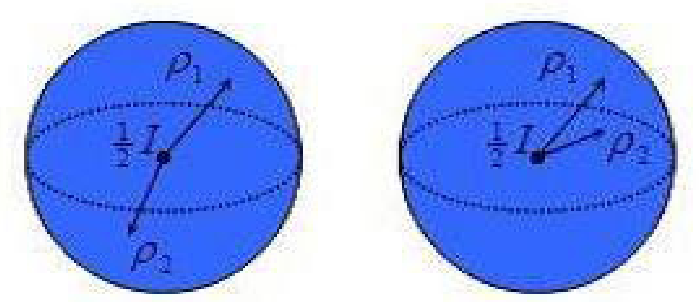}
    \caption{$\rho_1$ and $\rho_2$ are not commutative.}
    \label{fig: 2}
  \end{figure}

\begin{figure}
    \centering
    \includegraphics[width=0.35\textwidth]{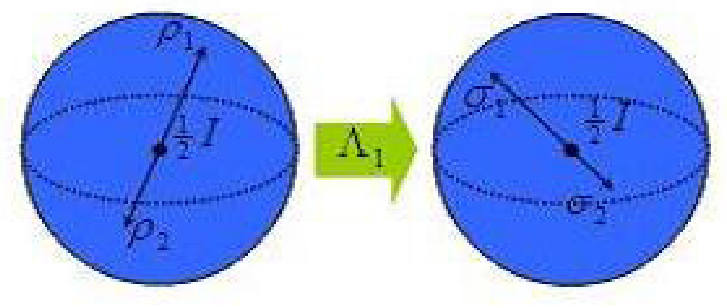}
   \includegraphics[width=0.35\textwidth]{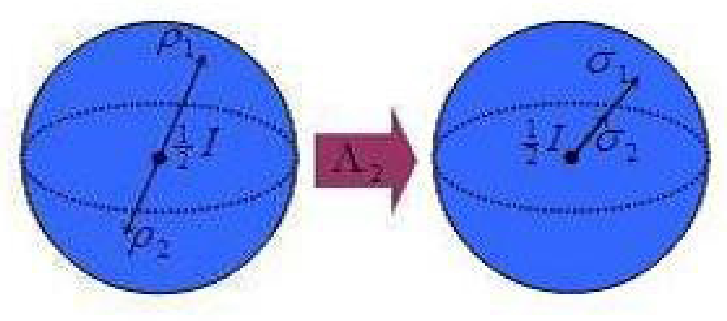}
    \caption{$\Lambda_{1}$ and $\Lambda_{2}$ are commutativity-preserving unital channels,
    they transforms commutative states to commutative ones.}
    \label{fig: 3}
  \end{figure}

At the end of this section, we give a geometric picture of commutative qubit states and commutativity-preserving qubit channels.
It is well-known that the \emph{Bloch ball}, whose boundary is the \emph{Bloch sphere}, corresponds to the space of all
two-level density matrices.
The surface represents all pure states while the interior of the Bloch sphere, the open Bloch ball, represents the mixed states. In particular, the center of the sphere corresponds to the maximally mixed state.
Indeed an arbitrary density matrix can be parameterized as
\begin{eqnarray*}
\rho
=\left(\begin{array}{cc} \frac{1}{2}+z& x-iy\\
 x+iy &\frac{1}{2}-z
\end{array}\right)
\end{eqnarray*}
with $x^2+y^2+z^2\leq\frac{1}{4}$.
It is customary to regard this an expansion in terms of the Pauli matrices
$\vec{\sigma}=(\sigma_x,\sigma_y,\sigma_z)$, so that
\begin{eqnarray}
\rho
=\frac{1}{2}I+\vec{r}\cdot\vec{\sigma}. \label{0829}
\end{eqnarray}
The vector $\vec{r}$ is known as the \emph{Bloch vector}.

\begin{figure}
    \centering
    \includegraphics[width=0.35\textwidth]{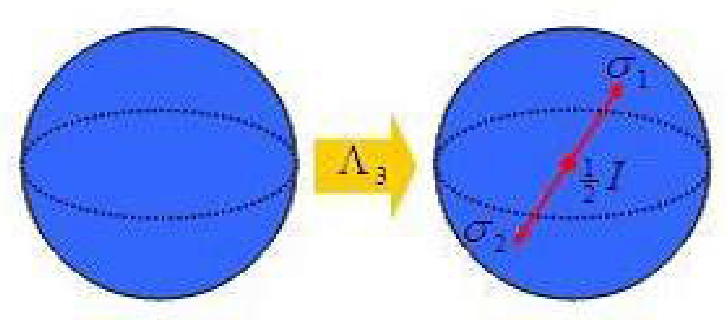}
    \includegraphics[width=0.35\textwidth]{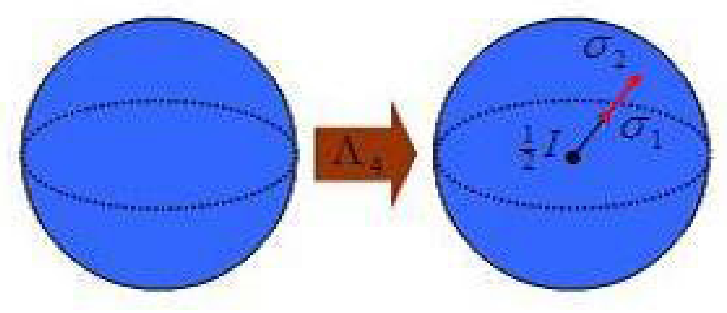}
    \caption{$\Lambda_{3}$ and $\Lambda_{4}$ are completely decohering channels,
    the red lines are the image of all qubit states under these channels.}
    \label{fig: 4}
  \end{figure}

\begin{figure}
    \centering
    \includegraphics[width=0.35\textwidth]{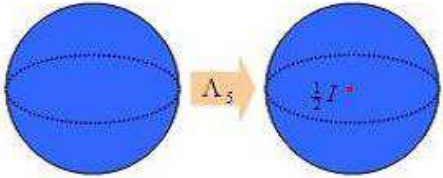}
    \caption{$\Lambda_5$ denotes the completely depolarizing channel which is a special case of completely decohering channel,
    it outputs the maximal mixed state for any input state.}
    \label{fig: 3}
  \end{figure}

Let \begin{eqnarray*}
\rho_k
=\left(\begin{array}{cc} \frac{1}{2}+z_k& x_k-iy_k\\
 x_k+iy_k &\frac{1}{2}-z_k
\end{array}\right),\quad k=1, 2.
\end{eqnarray*}
It is straightforward that $[\rho_1,\rho_2]=0$ if and only if
$x_1z_2=x_2z_1$, $y_1z_2=y_2z_1$ and $x_1y_2=x_2y_1$. Equivalently, $[\rho_1,\rho_2]=0$ if and only if
$\vec{r}_1=t\vec{r}_2$ for some real number $t$, where $\vec{r}_k$ denotes the Bloch vector of $\rho_k$, $k=1$, $2$.
\if
In fact, any hermitian matrix can be expressed as Eq.~(\ref{0829}) with $\vec{r}=(x,y,z)$ is an arbitrary
vector in $\mathbb{R}^3$ (namely, the condition $x^2+y^2+z^2\leq\frac{1}{4}$ is not necessarily satisfied). In such a case, we call $\vec{r}$ a generalized Bloch vector.
One can easily check that two hermitian matrices commute with each other if and only if the generalized Bloch vectors are proportionable, namely two hermitian matrices $A$, $B$ with $[A,B]=0$ if and only if
$\vec{r}_A=t\vec{r}_B$ for some real number $t$, where $\vec{r}_{A}$ (resp. $\vec{r}_{B}$) denotes the Bloch vector of $A$ (resp. $B$). That is
\fi

{\it Observation 1.}\quad Two qubit quantum states are commutative if and only if they are collinear with the center of the sphere in the Bloch ball (see fig. 1 and fig. 2).
\if
Generally, two hermitian matrices are commutative if and only if
they are collinear with the the origin of the coordinates in this sense.
\fi

{\it Observation 2.}\quad A qubit channel preserves commutativity in both directions if and only if it is a one-to-one transformation and it
maps line that collinear with the center of the ball to line that collinear with the center of the ball (see fig. 3).
A completely decohering qubit channel maps the Bloch ball to a line that collinear with the center of the ball (see fig. 4 and fig. 5).

Let $\rho$ be a qubit state with $\vec{r}_\rho=(x,y,z)$. Then a unitary evolution of $\rho$,
i.e., $U\rho U^\dag$ for some unitary matrix, corresponds to a rotation of the Bloch vector $\vec{r}_\rho$  around
the center of the ball. Write $\rho'=U\rho U^\dag$ with $\vec{r}_{\rho'}=(\tilde{x},\tilde{y},\tilde{z})$, then
$\tilde{x}^2+\tilde{y}^2+\tilde{z}^2=x^2+y^2+z^2$. Conversely, if $\rho_1$ and $\rho_2$ are two qubit state with $\vec{r}_{\rho_1}=(x_1,y_1,z_1)$, $\vec{r}_{\rho_2}=(x_2,y_2,z_2)$ and
$x_1^2+y_1^2+z_1^2=x_2^2+y_2^2+z_2^2$, then $\rho_1=U\rho_2 U^\dag$ for some unitary matrix $U$.
For simplicity, for a given state $\rho$, we denote by $\Pi_U\vec{r}_\rho$ the rotation of the Bloch vector $\vec{r}_\rho$, which is corresponding to the unitary evolution of the state $\rho$, $U\rho U^\dag$.

{\it Observation 3.}\quad Let $\Lambda$ be a qubit channel as in Eq.~(\ref{8}), and let
$\alpha=a+ib$, $\beta=c+id$ and $\gamma=e+if$, where $a$, $b$, $c$, $d$, $e$ and $f$ are real numbers,
$i$ is the imaginary unit. Then
\begin{eqnarray*}
\vec{r}_{\sigma}=\Pi_U(x',y',z'),\quad\forall\ \vec{r}_{\rho}=(x,y,z),
\end{eqnarray*}
where $\sigma=\Lambda(\rho)$, $x'=\lambda(xc-yd)+(1-\lambda)(xe-yf)$,
$y'=\lambda(cy+dx)-(1-\lambda)(ey+fx)$ and $z'=(2\lambda-1)z+2\sqrt{\lambda(1-\lambda)}(xa-yb)$.

\section{Local channels that nullify QD}

In the following let us turn to the question when a local channel
nullifies QD in any states.

Let $\Lambda$ be a channel acting on subsystem B. It is obvious that if
$\Lambda(\mathcal{B}(H_B))$ is commutative, then
$D_B((I_A\otimes\Lambda)\rho)=0$ for any state
$\rho\in\mathcal{S}(H_A\otimes H_B)$, namely, $\Lambda$ nullifies
QD in any states. Conversely, if $\Lambda$ nullifies QD in any
states, one can check that
 $\Lambda(\mathcal{B}(H_B))$ is commutative.
In fact, writing $\rho=\sum_{i,j}E_{ij}\otimes B_{ij}$ as in
Eq.~(\ref{l}),  $D_B((I_A\otimes\Lambda)\rho)=0$ yields that
$\Lambda(B_{ij})$'s are mutually commuting normal operators. \if
We may assume
that $\dim H_A\geq 3$ for simplicity. Then, for any
 $A,B\in{\mathcal B}(H_B)$, there exist positive numbers $a,b,c$ such that
 \begin{eqnarray*}
\begin{array}{rl}
\varrho_0
=&\frac{1}{(a+b+c){\rm Tr}(I_B)}(E_{11}\otimes P_1+E_{12}\otimes A\\
& +E_{21}\otimes A^\dag+E_{22}\otimes P_2
\end{array}
\end{eqnarray*}
is a state in ${\mathcal S}(H_A\otimes H_B)$.\fi
For any $A\in{\mathcal B}(H_B)$, there exist positive operators
$P_1,P_2,P_3,P_4$ such that
$A=P_1-P_2+i(P_3-P_4)$. Then for any $B\in{\mathcal B}(H_B)$,
\begin{eqnarray*}
\begin{array}{rl}
\varrho_0
=&\frac{1}{{\rm Tr}(P_1+P_2)}(E_{11}\otimes P_1+E_{12}\otimes B\\
& +E_{21}\otimes B^\dag+E_{22}\otimes P_2)
\end{array}
\end{eqnarray*}
and
\begin{eqnarray*}
\begin{array}{rl}
\sigma_0
=&\frac{1}{{\rm Tr}(P_3+P_4)}(E_{11}\otimes P_3+E_{12}\otimes B\\
& +E_{21}\otimes B^\dag+E_{22}\otimes P_4)
\end{array}
\end{eqnarray*}
are states in ${\mathcal S}(H_A\otimes H_B)$.
 It follows from
$D_B((I_A\otimes\Lambda)\varrho_0)=0$
and $D_B((I_A\otimes\Lambda)\sigma_0)=0$ that $\Lambda(A)$ and
$\Lambda(B)$ are commuting normal operators, which implies that
$\Lambda(\mathcal{B}(H_B))$ is commutative. Thus, the following
result is true.

{\it Theorem 3}.\quad Let $H_A$ and $H_B$ be complex Hilbert
spaces with $\dim H_A=m\geq2$ and $\dim H_B=n\geq2$, and let $\Lambda$ be a
channel acting on subsystem B. Then
 $D_B((I_A\otimes\Lambda)\rho)=0$ for any state
$\rho\in\mathcal{S}(H_A\otimes H_B)$ if and only if
$\Lambda$ is
a completely decohering channel.
\if
In particular, $\Lambda$ is
a completely decohering channel if and only if
it has the form as Eq.~(\ref{cc}).
 or $\Lambda(\cdot)=\frac{1}{n}{\rm Tr}(\cdot)I_B$.
 There exists an $n-$outcome POVM for system B, i.e.,
$\{W_i\}_{i=1}^n\subset{\mathcal B}(H_B)$ with
$\sum_iW_i=I_B$, $W_i\geq0$, and an orthonormal basis  $\{|e_i\rangle\}$ of
$H_B$ such that
\begin{eqnarray}
\Lambda(A)=\sum\limits_i{\rm Tr}(W_iA)|e_i\rangle\langle e_i|
\end{eqnarray}
holds for all $A\in{\mathcal B}(H_B)$.
\fi

Finally, let us discuss the form of completely decohering channel.

{\it Theorem 4.}\quad Let $\Lambda$ be a channel acting on the system associated with a $n-$dimensional complex Hilbert space $H$, $n\geq2$. Then $\Lambda$ is a completely decohering channel if and only if
there exist an $n-$outcome POVM for system B, i.e.,
$\{W_i\}_{i=1}^n\subset{\mathcal
B}(H)$ with $\sum_iW_i=I$, $W_i\geq0$, $i=1$, 2, $\dots$, and an
orthonormal basis, $\{|e_i\rangle\}$, of $H$, such that
\begin{eqnarray}
\Lambda(A)=\sum\limits_i^{n}{\rm Tr}(W_iA)|e_i\rangle\langle e_i|\
\mbox{ for all }\ A. \label{cc}
\end{eqnarray}

{\it Proof.}\quad The `if part' is clear.
We show the `only if' part.
By definition, $\Lambda$ is completely decohering implies that
 $\Lambda(\mathcal{B}(H))$ is commutative. Then $\Lambda(A)$ is
normal for any $A\in\mathcal{B}(H)$ since
$[\Lambda(A),\Lambda(A^\dag)]=[\Lambda(A),\Lambda(A)^\dag]=0$. Hence
all elements in $\Lambda(\mathcal{B}(H))$ are normal and mutually
commutative. It follows that there exist positive linear functionals
$f_i$ of ${\mathcal B}(H)$, $i=1$, 2, $\dots$, $n$ and an
orthonormal basis $\{|e_i\rangle\}$ of $H$ such that
\begin{eqnarray*}
\Lambda(A)=\sum\limits_if_i(A)|e_i\rangle\langle e_i|.
\end{eqnarray*}
Therefore there exist positive operators $W_i\in {\mathcal B}(H)$,
$i=1$, 2, $\dots$, $n$, so that $f_i(A)={\rm Tr}(W_iA)$ holds for
any $A\in\mathcal{B}(H)$. Since $\Lambda$ is a trace-preserving
map we have ${\rm Tr}(A)=\sum_i{\rm Tr}(W_iA)={\rm
Tr}((\sum_iW_i)A)$ holds for any $A\in\mathcal{B}(H)$, which leads
to
 $\sum_iW_i=I_B$. That is $\Lambda$ has the form as desired.\hfill$\blacksquare$

\section{Conclusions}

By the feature of commutativity-preserving
linear map, we got a clear picture of
local channels that cannot create the quantum discord.
We obtained an exact form of local channel that cannot
create QD for zero QD states.
\if
These channels are divided into two classes:
the completely decohering channel and the
nontrivial isotropic channel when the subsystem is of $n$-dimensional with $n\geq3$
(is the commutativity-preserving unital channel when the subsystem is a qubit system).\fi
Consequently,
the conjecture in Ref.~\cite{Hu} was confirmed and
the Theorem 1 in Ref.~\cite{Streltsov} was improved.
\if
Moreover, we showed that the former one nullifies QD
in any state and the later one preserves zero QD
states in both directions when $n\geq3$.
\fi
We also found that, remarkably, the qubit case is
quite different from the higher dimensional case since
there exist qubit local channels that are not
isotropic channels while they preserves zero QD states
in both directions as well.
In addition, the geometric picture of the commutative qubit states in the Bloch ball was depicted.
We hope that our results would be useful in
realizing quantum communication and quantum computation experimentally.


Our results lead to interesting questions for further study:
what is the form of local channel
$\Lambda_{a/b}:\mathcal{B}(H_{A/B})\rightarrow\mathcal{B}(H_{A/B})$
if $D_{A/B}(\rho)=D_{A/B}((\Lambda_a\otimes\Lambda_b)\rho)$?
\if In addition, by Proposition 1, we know that $\Lambda_{a/b}$
admits the form in item (3) of Proposition 1.\fi
We conjecture that
$\Lambda_{a/b}$ is the unitary operation in such a case,
however the proof may be difficult since the calculation of QD is a
hard work in general.
Moreover,
what is the form of a total channel
$\Lambda:\mathcal{B}(H_A\otimes H_B)\rightarrow\mathcal{B}(H_A\otimes H_B)$
if it satisfies one of the following conditions:
(1) $D_{A/B}(\rho)=0\Rightarrow D_{A/B}(\Lambda(\rho))=0$
(or $D_{A/B}(\rho)=0\Leftrightarrow D_{A/B}(\Lambda(\rho))=0$);
(2) $D_{A/B}(\rho)=D_{A/B}(\Lambda(\rho))$?

\begin{acknowledgments}
\if The authors gratefully acknowledge Pro.
Shengjun Wu and Dr. Xueyuan Hu for a critical reading of the manuscript
and their helpful suggestions. \fi
This work is partially supported by
Natural Science Foundation of China
(11171249,11101250) and Research start-up fund for Doctors
of Shanxi Datong University (2011-B-01).
We thank Karol \.{Z}yczkowski and Yiu-Tung Poon for valuable discussions.
\end{acknowledgments}

\section*{Appendix: Commutativity-preserving linear maps on $\mathcal{M}_2$}

In order to prove Theorem 2, the theorem below is necessary.

{\it Theorem 5}.\quad Let $\phi:\mathcal{M}_2\rightarrow \mathcal{M}_2$
be a hermitian-preserving linear map. Then $\phi$ preserves
commutativity if and only if (a) either there exist two hermitian matrices
$W_1$ and $W_2$, and an orthonormal basis $\{|e_1\rangle,
e_2\rangle\}$ of $\mathbb{C}^2$, such that
\begin{eqnarray*}
\phi(A)={\rm Tr}(W_1A)|e_1\rangle\langle e_1|
+{\rm Tr}(W_2A)|e_2\rangle\langle e_2|, \label{ddd}
\end{eqnarray*}
or (b) there exist a unitary matrix $U$, and real numbers
$\lambda_i$, $\mu_i$ with $\lambda_1+\mu_1=\lambda_2+\mu_2$ such
that
\begin{eqnarray}
\phi(E_{11})=
U\left(\begin{array}{cc}
\lambda_1&0\\
0&\lambda_2
\end{array}
\right)U^\dag, \label{dd}
\end{eqnarray}
\begin{eqnarray}
\phi(E_{22})=
U\left(\begin{array}{cc}
\mu_1&0\\
0&\mu_2
\end{array}
\right)U^\dag, \label{d}
\end{eqnarray}
where $E_{ij}$ denotes the 2 by 2 matrix with $(i,j)-$entry 1 and
others 0. \if we have
\begin{eqnarray}
\begin{array}{rcl}
\phi(A)&=&
U\left(\begin{array}{cc}
a_{11}\lambda_1+a_{22}\mu_1&0\\
0&a_{11}\lambda_2+a_{22}\mu_2
\end{array}
\right)U^\dag\\
&&+a_{12}X+a_{21}X^\dag \label{ddd}
\end{array}
\end{eqnarray}
holds for all $A=[a_{ij}]\in\mathcal{M}_2$.\fi

{\it Proof}.\quad Suppose that $\phi(I)\neq0$. Or else, replace
$\phi$ by $\psi$ with $\psi(A)=\phi(A)+{\rm Tr}(A)I$.

There are two different cases: (1) $\phi(I)$ is not a scalar matrix,
and (2) $\phi(I)=\lambda I$ for some $\lambda\neq 0$.

If $\phi(I)$ is not a scalar matrix,
there are two different subcases:

{\it Case 1}\quad  ${\rm rank} \phi(I)=2$. In this case, since
$[\phi(X),\phi(I)]=0$ holds for all $X\in\mathcal{M}_2$, there exist
two hermitian matrices $W_1$ and $W_2$, and an orthonormal basis,
$\{|e_1\rangle, e_2\rangle\}$ of $\mathbb{C}^2$, such that
\begin{eqnarray}
\phi(A)={\rm Tr}(W_1A)|e_1\rangle\langle e_1|+{\rm
Tr}(W_2A)|e_2\rangle\langle e_2|. \label{o}
\end{eqnarray}
Consequently, $\phi(\mathcal{M}_2)$ is commutative.

{\it Case 2}\quad  ${\rm rank} \phi(I)=1$. We may assume that
$\phi(I)=\gamma|\xi\rangle\langle\xi|$ for some unit vector
$|\xi\rangle$ and real number $\gamma\neq0$. Let $|\zeta\rangle$ be
a unit vector that is orthogonal to $|\xi\rangle$. Then
\begin{eqnarray*}
\phi(\mathcal{M}_2)\subseteq\{x|\xi\rangle\langle\xi|+y|\zeta\rangle\langle\zeta|:
x,y\in\mathbb{C}\}.
\end{eqnarray*}
Since $\phi$ maps a hermitian matrix into a hermitian one and any matrix
is a linear combination of hermitian matrices, we can thus conclude
that $\phi$ still has the form as in Eq.~(\ref{o}).

Conversely, if $\phi$ has the form as in Eq.~(\ref{o}),
then it is clear that $\phi$ preserves commutativity.

We now suppose that $\phi(I)=\lambda I$ for some $\lambda\neq 0$.
Since $[E_{11},E_{22}]=0$, we have $[\phi(E_{11}),\phi(E_{22})]=0$.
Thus there exists a $2$ by $2$ unitary matrix $U$ such that
 Eqs.~(\ref{dd}) and (\ref{d}) hold. Moreover,
 $\lambda_1+\mu_1=\lambda=\lambda_2+\mu_2$.

On the other hand, assume that $\phi$ satisfies
Eqs.~(\ref{dd}) and (\ref{d}) with $\lambda_1+\mu_1=\lambda_2+\mu_2$. Take
real numbers $x,z, \alpha, \gamma$ and complex numbers $y,\beta$ so
that
$\phi(E_{12}+E_{21})=U\left(\begin{array}{cc}
x&y\\
\bar{y}&z
\end{array}
\right)U^\dag$ and
$\phi(iE_{12}-iE_{21})=U\left(\begin{array}{cc}
\alpha&\beta\\
\bar{\beta}&\gamma
\end{array}
\right)U^\dag$.
Then
\begin{widetext}
\begin{eqnarray*} 
\phi\left(\left(\begin{array}{cc}
a_{11}&a_{12}\\
a_{21}&a_{22}
\end{array}
\right)\right)
= U\left[\left(\begin{array}{cc}
\lambda_1a_{11}+\mu_1 a_{22}&0\\
0&\lambda_2a_{11}+\mu_2a_{22}
\end{array}
\right)
+a_{12}X+a_{21}X^\dag \right]U^\dag
\end{eqnarray*}
holds for any matrix $\left(\begin{array}{cc}
a_{11}&a_{12}\\
a_{21}&a_{22}
\end{array}
\right)$, where
$X=U^\dag\phi(E_{12})U=
 \frac{1}{2}\left(\begin{array}{cc}
x&y\\
\bar{y}&z
\end{array}
\right)-\frac{i}{2}\left(\begin{array}{cc}
\alpha &\beta\\
\bar{\beta}&\gamma
\end{array}
\right)$.
In particular,
\begin{eqnarray*}
\phi\left(\left(\begin{array}{cc}
a&u+iv\\
u-iv&c
\end{array}
\right)\right)
=U\left(\begin{array}{cc} a\lambda_1+c\mu_1+ux+v\alpha &uy+v\beta \\
u\bar{y}+v\bar{\beta} &a\lambda_2+c\mu_2+uz+v\gamma
\end{array}
\right)U^\dag
\end{eqnarray*}
\end{widetext}
for any hermitian matrix $\left(\begin{array}{cc}
a&u+iv\\
u-iv&c
\end{array}
\right)$. Note that, two hermitian matrices
\begin{eqnarray*}
A=\left(\begin{array}{cc}
a&u_1+iv_1\\
u_1-iv_1&c
\end{array}
\right), B=\left(\begin{array}{cc}
d&u_2+iv_2\\
u_2-iv_2&f
\end{array}
\right),
\end{eqnarray*}
 are commuting if and only if
$(a-c)u_2=u_1(d-f)$, $(a-c)v_2=v_1(d-f)$ and $u_1v_2=u_2v_1$.

Then, by the above fact and noticing that
$\lambda_1-\lambda_2=\mu_2-\mu_1$,  one can check that, for any
hermitian matrices $A,B$, $[A,B]=0\Rightarrow [\phi(A),\phi(B)]=0$.
So $\phi$ preserves commutativity for hermitian matrices. Then by
\cite[Corollary 1]{Choi87}, we know that $\phi$ preserves
commutativity. \hfill$\blacksquare$

\nocite{*}

\bibliography{apssamp}

\end{document}